%% file: OMIT_Submit.tex
\documentclass[aps,prl,superscriptaddress,twocolumn]{revtex4-1}
\pdfoutput=1
\usepackage[version=3]{mhchem} 
\usepackage[dvips]{epsfig}
\usepackage{graphicx}  
\usepackage{dcolumn}   
\usepackage{bm}        
\usepackage{amssymb}   
\usepackage{color}
\usepackage{float}
\usepackage{hyperref}
\hypersetup{
     colorlinks   = true,
     citecolor    = blue,
     linkcolor    = red,
     urlcolor     = black
     }
\usepackage{etoolbox}
\patchcmd{\section}
  {\centering}
  {\raggedright}
  {}
  {}
\patchcmd{\subsection}
  {\centering}
  {\raggedright}
  {}
  {}

\begin{document}

\title{Optomechanically induced transparency and cooling in thermally stable diamond microcavities}

\author{David P.  Lake}
\affiliation{Contributed equally to this work}
\affiliation{Department of Physics and Astronomy and Institute for Quantum Science and Technology, University of Calgary, Calgary, AB, T2N 1N4, Canada}
\affiliation{National Institute for Nanotechnology, Edmonton, AB, T6G 2M9, Canada}

\author{Matthew Mitchell}
\affiliation{Contributed equally to this work}
\affiliation{Department of Physics and Astronomy and Institute for Quantum Science and Technology, University of Calgary, Calgary, AB, T2N 1N4, Canada}
\affiliation{National Institute for Nanotechnology, Edmonton, AB, T6G 2M9, Canada}

\author{Yasmeen Kamaliddin}
\affiliation{Department of Physics and Astronomy and Institute for Quantum Science and Technology, University of Calgary, Calgary, AB, T2N 1N4, Canada}

\author{Paul E. Barclay}
\email{pbarclay@ucalgary.ca}
\affiliation{Department of Physics and Astronomy and Institute for Quantum Science and Technology, University of Calgary, Calgary, AB, T2N 1N4, Canada}
\affiliation{National Institute for Nanotechnology, Edmonton, AB, T6G 2M9, Canada}

\begin{abstract}
Diamond cavity optomechanical devices hold great promise for quantum technology based on coherent coupling between photons, phonons and spins. These devices benefit from the exceptional physical properties of diamond, including its low mechanical dissipation and optical absorption. However the nanoscale dimensions and mechanical isolation of these devices can make them susceptible to thermo-optic instability when operating at the high intracavity field strengths needed to realize coherent photon--phonon coupling. In this work, we overcome these effects through engineering of the device geometry, enabling operation with large photon numbers in a previously thermally unstable regime of red-detuning. We demonstrate optomechanically induced transparency with cooperativity $ > 1$ and normal mode cooling from 300 K to 60 K, and predict that these device will enable coherent optomechanical manipulation of diamond spin systems.
\end{abstract}

\maketitle

\section{Introduction}

Nanophotonic cavity optomechanical devices localize light within nanostructures supporting both optical and mechanical resonances,  creating large optical forces that can coherently couple light to phonons of a mechanical mode. These devices provide a testbed for fundamental studies of quantum science \cite{ref:aspelmeyer2014co, ref:treutlein2014hms}, with hallmark experiments demonstrating phenomena such as optomechanically induced transparency \cite{ref:weis2010oit, ref:safavi2011eit, ref:liu2013eit}, optomechanical cooling \cite{ref:chan2011lcn}, observation of a mechanical resonator's zero point motion \cite{ref:safavinaeini2012oqm}, quantum optical--mechanical correlations \cite{ref:cohen2015pci, ref:riedinger2016ncc,ref:purdy2017qcrt, ref:sudhir2017qco, ref:nielsen2017mos}, and entanglement between mechanical resonators \cite{ref:riedinger2017rqe}. Within the realm of quantum technology, the ability of these systems to coherently interface GHz frequency phonons with optical photons has sparked efforts to create  transducers \cite{ref:regal2011cec} that optomechanically convert quantum information between photonic channels and solid state  \cite{ref:lee2017trs} or superconducting microwave \cite{ref:bochmann2013ncb, ref:fong2014mac, ref:pitanti2015soe, ref:balram2016ccr} qubits via a shared mechanical coupling.

Diamond cavity optomechanical devices \cite{ref:mitchell2016scd, ref:burek2016doc} are poised to advance experiments in quantum optomechanics, in part thanks to device performance improvements from diamond's  best-in-class Young's modulus, low intrinsic mechanical dissipation,  high thermal conductivity, and large optical transparency window from $\sim 230~\text{nm}$ to far--IR \cite{ref:aharonovich2011dp}. In addition, these devices  offer an interface between highly coherent diamond colour centre spins and optically controlled phonons via strain coupling \cite{ref:lee2017trs}, which could enable quantum transducers between spins and quantum phononic and photonic states \cite{ref:stannigel2012oqi, ref:stannigel2010otl, ref:didier2014qtc}, as well as platforms for entangling remote spins via nanomechanical coupling \cite{ref:bennett2013pis, ref:albrecht2013cnvp}. To date, nanomechanical devices used for spin manipulation have relied on piezo actuated phonon-strain coupling  \cite{ref:macquarrie2013msc, ref:ovartchaiyapong2014dsc, ref:teissier2014scn,  ref:meesala2016esc, ref:momenzadeh2016tcd, ref:golter2016oqc}. Incorporating coherent cavity optomechanics would enable optical control of phonon-spin interactions with sensitivity necessary for operation at the single phonon level and enable new photon-spin interfaces that are independent of the spin optical properties.

A requirement for realizing coherent photon--phonon interactions is an optomechanical cooperativity $C \equiv 4 N g_0^2 / \kappa_\text{o}\Gamma_\text{m} > 1$, where $N$ is the intracavity photon number, $g_0$ is the single--photon optomechanical coupling rate, and $\kappa_\text{o}$ and $\Gamma_\text{m}$ are the optical cavity and mechanical resonator energy decay rates, respectively \cite{ref:aspelmeyer2014co}. We have developed single--crystal diamond microdisks that achieve $C > 1$ by coupling optical whispering gallery modes to GHz frequency ($\omega_\text{m}$) mechanical radial breathing mode resonances \cite{ref:mitchell2016scd}. The microdisk mechanical resonances have a large mechanical quality factor $Q_\text{m} \equiv \omega_\text{m}/\Gamma_\text{m}$ that is enhanced by restriction of phonon leakage into the substrate by a nanoscale pedestal, while the optical modes have optical quality factor $Q_o \equiv \omega_\text{o}/\kappa_\text{o} \sim 10^5$, where $\omega_\text{o}/2\pi \sim 200~\text{THz}$ is the optical mode frequency, and can support $N > 10^6$ photons without suffering from nonlinear absorption that degrades device performance in other less transparent materials such as silicon \cite{ref:barclay2005nrs}.

However, in previous work, at high $N$ the device performance becomes limited by linear optical absorption and accompanying heating. The resulting change in microdisk temperature is exacerbated by the $\sim100~\text{nm}$ waist of the microdisk pedestal, in which thermal conductivity is reduced at room temperature due to size effects by approximately an order of magnitude compared to in bulk diamond \cite{ref:mitchell2016scd}, resulting in thermo-optic instability \cite{ref:carmon2004dtb} for red laser--cavity detunings needed for applications such as coherent phonon--photon coupling and  optomechanical cooling.
Here we overcome this limitation through modification of the microdisk pedestal shape to improve its thermal conductivity without affecting $Q_\text{m}$, enabling demonstration of optomechanically induced transparency, a hallmark of coherent phonon-photon coupling, as well as stable optomechanical cooling in diamond microdisks for the first time.

\begin{figure}[h]
\epsfig{figure=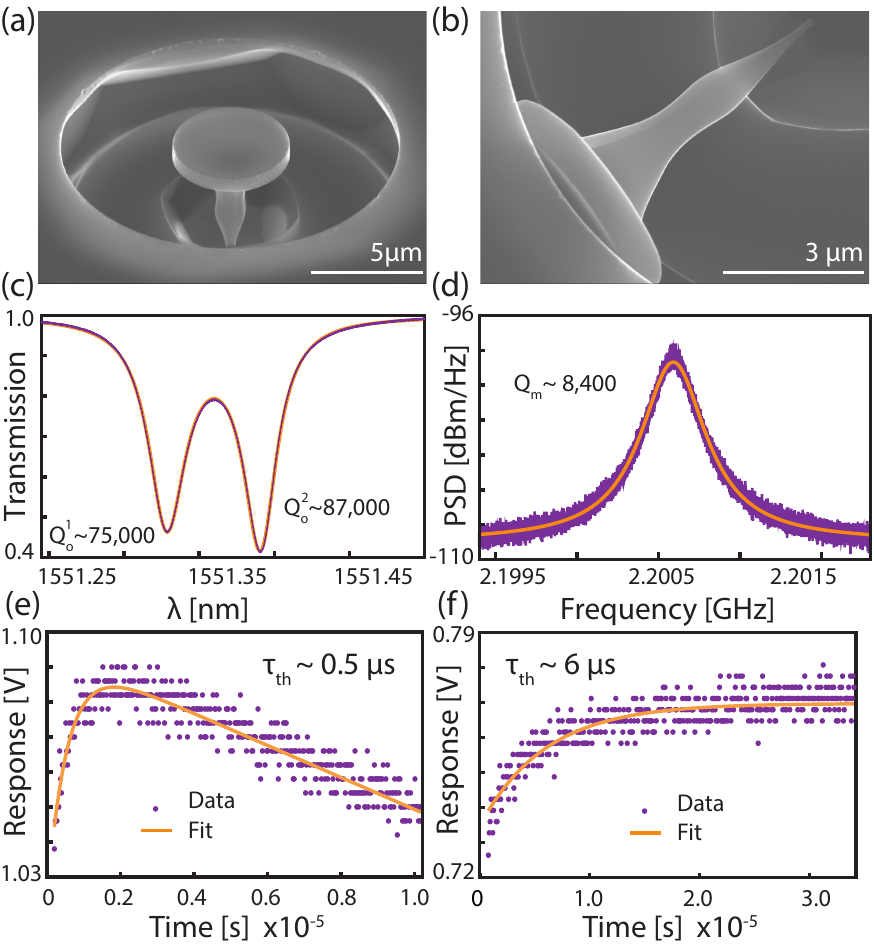, width=1\linewidth}
 \caption{Characterization of single--crystal diamond microdisk structures fabricated using modified plasma undercutting method. (a) Scanning electron micrograph (SEM) of a ``flared'' microdisk structure. (b) Top down SEM view of the modified pedestal shape using the modified undercutting method described above. (c) Typical high--$Q_\text{o}$ doublet optical mode for the device studied here. (d) RBM of microdisk measured at low input power exhibiting a $Q_\text{m} \sim 8,400$. (e,f) Response of the optical transmission for an input step function for microdisks with similar sized flared and hourglass shaped pedestals, respectively. The exponential rise is fit to extract the thermal time constant. The negatively sloped feature in (e) is an EDFA artifact, resultant from the large input powers required to achieve thermal bistability.}
\label{fig:SEM}
\end{figure}

\section{Device characterization: optical, mechanical and thermal properties}

An example of the diamond microdisk devices studied here is shown in the scanning electron micrograph image in Fig.\ \ref{fig:SEM}(a). These devices were fabricated by modifying the diamond undercutting process demonstrated in Ref.\ \cite{ref:mitchell2016scd}, as described in Supplement 1, Section 1,  to shape the pedestal supporting the microdisk into the flared profile shown in Fig.\ \ref{fig:SEM}(b). Note that the device in Fig.\ \ref{fig:SEM}(b) was broken as result of being over--undercut, and its minimum pedestal dimension is smaller than that of the devices studies here, which are $\sim 350$ nm. The optical and mechanical modes of the microdisks were characterized by monitoring the transmission of a tuneable diode laser (Newport TLB 6700B) through a dimpled optical fiber taper evanescently coupled to the microdisk.   Scans of transmission for varying laser wavelength ($\lambda_p$) when coupled to the $5~\mu\text{m}$ diameter microdisk considered in the remainder of this work revealed resonances such as those shown in Fig.\ \ref{fig:SEM}(c). The doublet resonance structure indicates the presence of standing wave optical modes created by backscattering in the microdisk. The modes studied here have a central wavelength of $\lambda_\text{o}\equiv 2\pi\text{c}/\omega_\text{o} \sim 1550$ nm, splitting of $\frac{\lambda_\text{o}^2}{2\pi c}\kappa_\text{bs} \sim 70$ pm, where $\kappa_\text{bs}$ is the backscattering rate, and intrinsic (unloaded)  $Q_\text{o}^{(1,2)} = \omega_\text{o}/\kappa_\text{o}^{(1,2)} = 7.4\times 10^4$ and $8.7\times 10^4$, where $\kappa_\text{o}^{(1,2)}$ are the optical energy decay rates of the blue and red shifted modes of the doublet, respectively.

The microdisk's mechanical resonances were probed by fixing $\lambda_p$ slightly off-resonance from the cavity modes and monitoring fluctuations in optical transmission due to mechanical motion using a high-speed photodetector (Newport 1554-B). Initial measurements were performed at low optical input power $P_\text{in}$ to avoid modifying the mechanical mode dynamics via optomechanical back action  \cite{ref:aspelmeyer2014co}. The power spectral density (PSD) of this signal, as analyzed on a real time spectrum analyzer (Tektronix RSA5106A) and shown in Fig.\ \ref{fig:SEM}(d), reveals the thermal motion of a mechanical resonance at $\omega_\text{m}/2\pi \sim 2.2\ \text{GHz}$, with a mechanical quality factor $\mathrm{Q_m} = \omega_\text{m}/\Gamma_\text{m} \sim 8,400$. Comparison to COMSOL finite element simulations of the microdisk suggest that this mechanical mode is the fundamental radial breathing mode (RBM) of the microdisk, as the measured $\omega_\text{m}$ is within $5\%$ of the simulated value. The RBM studied here has an effective mass predicted from simulation of $m_\text{eff}\sim$ 45 pg, corresponding to a quantum zero point motion amplitude, $x_\text{zpm} \sim 0.30$ fm, where $x_\text{zpm} = \sqrt{\hbar/2m_\text{eff}\omega_\text{m}}$.

Reaching the regime of coherent optomechanical coupling can in principle always be achieved by operating with high enough $P_\text{in}$ to increase $N$ so that $C > 1$. In practice, even in absence of nonlinear absorption, $N$ is limited by linear absorption and thermo-optic dispersion, particularly in small optical mode volume devices such as microdisks.   The microdisk pedestal shape plays a critical role in determining whether $C >1$ can be reached, as it influences $Q_\text{m}$ ($\propto C$) \cite{ref:mitchell2016scd} as well as the device's ability to conduct thermal energy away from the microdisk and mitigate optical heating. The importance of the pedestal's thermal conductance can be seen by considering the threshold for $N$ above which the microdisk becomes bistable  due to the thermo-optic effect \cite{ref:gibbs1985obc, ref:almeida2004obs}, for $\lambda_\text{s}$ red-detuned from $\lambda_\text{o}$ as required for optomechanically induced transparency \cite{ref:weis2010oit, ref:safavi2011eit, ref:liu2013eit} and cooling \cite{ref:chan2011lcn}:
\begin{equation}
    \left(\frac{\eta P_\text{in} Q_\text{o}}{\omega_\text{o}}\right) \equiv N < \frac{C_p}{|\beta| \tau_\text{th}\omega_\text{o}}\left(\frac{ Q_\text{abs}}{Q_\text{o}}\right)
\label{eqn:bistable}
\end{equation}
\noindent (see Supplement 1, Section 2, is). Here $\tau_\text{th}$ and $C_p$ are the microdisk thermal time constant and the heat capacity, respectively, and $\beta = d\lambda_\text{o}/dT$ is the microdisk thermo-optic coefficient that accounts for thermal expansion and refractive index temperature dependence \cite{ref:carmon2004dtb}. The on-resonance fiber taper waveguide-microdisk power coupling efficiency is defined as $\eta$, and $1/Q_\text{abs}$ is the contribution to $1/Q_\text{o}$ from linear absorption. Equation \eqref{eqn:bistable} illustrates the inverse relationship between $\tau_\text{th}$ and maximum $N$. Note that in the anomalous case of $\beta < 0$, the cavity becomes bistable for blue instead of red detuning.

To determine the impact of the pedestal shape on the properties of the microdisks, we compare the flared pedestal devices from this work with  hourglass pedestal devices studied previously \cite{ref:mitchell2016scd}. In Ref.\  \cite{ref:mitchell2016scd} it was found that increasing pedestal width degraded $Q_\text{m}$. However, the $Q_\text{m}$ of the flared pedestal device measured here is similar to the best valued reported for the hourglass microdisks, despite the larger width of the flared pedestals. However, as shown by measurements of $\tau_\text{th}$ in  Fig.\ \ref{fig:SEM}(e-f), the highest $Q_\text{m}$ flared and hourglass pedestal microdisks have $\tau_\text{th} \sim 0.5 \ \mu$s and $6\ \mu$s, respectively, indicating that the flared pedestal devices studied here can support over an order of magnitude larger $N$ in the red-detuned regime compared to previous hourglass pedestal devices.   Here $\tau_\text{th}$ was measured by monitoring the response of the microdisk to an optical pulse that causes $\lambda_p$ to shift via the photothermal effect, as described in Supplement 1, Section 2.

\section{Optomechanical spring effect}

\begin{figure}[h]
\epsfig{figure=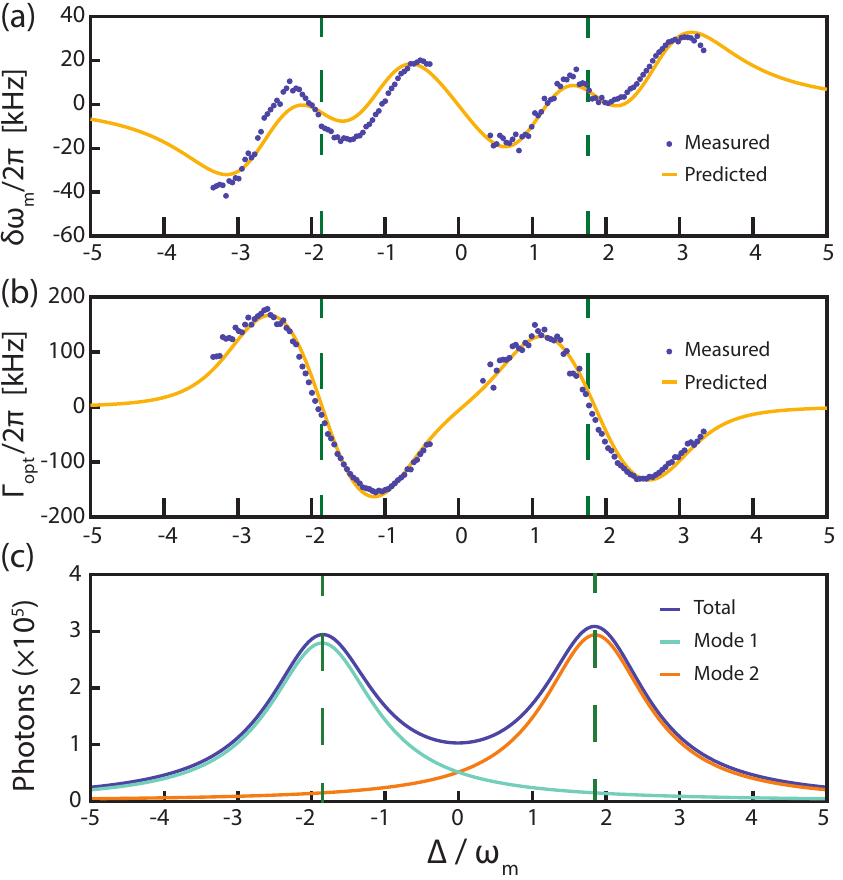, width=1\linewidth}
 \caption{Characterization of optomechanical coupling via the spring effect for the device shown in Fig.\ \ref{fig:SEM}. Shift in (a)  mechanical resonance frequency $\delta\omega_\text{m}$ and (b) optomechanical damping $\Gamma_\text{opt}$  as a function of laser detuning. The fits are from the optomechanical spring effect calculation with a single--photon optomechanical coupling rate of $g_0/2\pi \sim 18$ kHz as the sole free parameter, and the measured $N$ shown in (c).  Dashed lines indicate  $\Delta \pm \kappa_\text{bs}/2 = 0$ (corresponding to the resonance frequency of each doublet mode), and  illustrate that $\delta\omega_\text{m}$ and $\Gamma_\text{opt} = 0$ when the laser is on-resonance with the cavity mode.}
\label{fig:Spring}
\end{figure}

The  optomechanical parameters of the system were further probed by means of the optical spring effect \cite{ref:aspelmeyer2014co, ref:braginsky1967pee, ref:braginsky1970idp}. For this measurement, $P_\text{in}$ was increased via an erbium doped fiber amplifier (EDFA: Pritel LNHPFA-30) connected to the tunable laser output. The laser wavelength was then discretely stepped across the optical cavity resonances, and the PSD of the transmitted signal was acquired at each step. By fitting a Lorentzian lineshape to the PSD, both $\omega_\text{m}(\Delta; P_\text{in}) = \omega_m^0 + \delta\omega_\text{m}(\Delta; P_\text{in})$ and $\Gamma_\text{m}(\Delta;  P_\text{in}) = \Gamma_\text{m}^0 + \Gamma_\text{opt}(\Delta; P_\text{in})$ as a function of pump-cavity detuning, $\Delta = \omega_\text{p} - \omega_\text{o}$, were extracted. The results of this measurement for intermediate input power  ($P_\text{in} \sim 4.7$ mW) are shown in Figs.\ \ref{fig:Spring}(a) and \ref{fig:Spring}(b). Here the data is fit to analytic expressions for the predicted values of the optomechanical spring effect $\delta\omega_\text{m}$ and optomechanical damping $\Gamma_\text{opt}$, taking into account the doublet nature of the optical mode  (see Supplement 1, Section 3). Using measurements of $N(\Delta)$ shown in Fig.\ \ref{fig:Spring}(c) determined from the power dropped into the microdisk and $\kappa$ we are able to extract the sole fitting parameter, $g_0/2\pi \sim 18$ kHz. Unlike previous measurements of the optical spring effect in diamond microdisks \cite{ref:mitchell2016scd},  $\delta\omega_\text{m}$ here was dominated by optomechanical back-action owing to the device's reduced $\tau_\text{th}$ and low optical heating.

The cavity optomechanical damping, $\Gamma_\text{opt}$, modifies the mechanical normal-mode temperature, $T_\text{eff}$, as in experiments of ground state cooling \cite{ref:chan2011lcn}, or generation of self--oscillations that drive stress fields for coupling to diamond colour center spins \cite{ref:mitchell2016scd}. The normal-mode temperature can be measured as a function of $\Delta$ from the area under the PSD normalized by the wavelength dependent optomechanical transduction (see Supplement 1, Section 3) and assuming that for large $|\Delta|$ the cavity is in thermal equilibrium with the room-temperature environment (see Supplement 1, Section 4). Figure  \ref{fig:Gamma}(a) shows the  normal mode temperature $T_\text{eff}$ and corresponding phonon occupation $n_\text{m}$ measured using this technique. Also shown is a prediction of $T_\text{eff}$ obtained by inputting the fit of $\Gamma_\text{opt}(\Delta)$ from Fig.\ \ref{fig:Spring}(b) to the optomechanical back-action cooling expression,
\begin{equation}\label{eq:phonon_n}
 n_\text{m}  =  n_\mathrm{th}  \times\frac{ \Gamma_\text{m}^0}{\Gamma_\text{m}^0+\Gamma_\mathrm{opt}(\Delta)},
\end{equation}
where $n_\text{m}$ is the final phonon number, $n_\mathrm{th} = k_B T/\hbar\omega_\text{m}$ is the equilibrium thermal phonon occupation \cite{ref:aspelmeyer2014co}. This expression is valid in the high--$T$ limit $n_\text{th} \gg n_\text{min}$ applicable here, where $n_\text{min} = 0.088$ is the minimum backaction limited phonon number achievable through optomechanical damping. The good agreement indicates that the $\Delta$ dependent normalization of the PSD is accurate, and that optical absorption and heating is small compared with changes to $n_\text{m}$ from optomechanical backaction.

At higher $P_\text{in}$, microdisk heating and modal thermo-optic dispersion become significant, and the transduction calibration could not be readily applied to measurements of  PSD area for varying $\Delta$. However, measurement of optomechanical cooling using Eq.\ \eqref{eq:phonon_n} with $\Delta$ optimized to maximize $\Gamma_\text{opt}$ was possible: for $N\sim 1\times10^6$ ($P_\text{in} \sim 20 $ mW), $T_\text{eff} = 60~\text{K}$  ($n_\text{m} = 588$ phonons) was measured, as shown in Fig.\ \ref{fig:Gamma}(b). Here $T_\text{eff}$ includes an increase in bath temperature (i.e.\ $n_\text{th}$) of  $4~\text{K}$ due to optical heating, inferred from the  shift in $\lambda_\text{o}$ calibrated by its independently measured temperature dependence (Supplement 1, Section 2).  This optimized cooling was obtained when red-detuned by $\omega_\text{m}$ from the higher-$Q_\text{o}$ doublet mode, as expected for a sideband-resolved cavity optomechanical device. This detuning was not achievable in previous work with hourglass pedestal microdisks due to an inability to operate at red-detuning with $P_\text{in}$ large enough to significantly reduce $n_\text{m}$ because of thermal instability \cite{ref:mitchell2016scd}.

\begin{figure}[ht]
\epsfig{figure=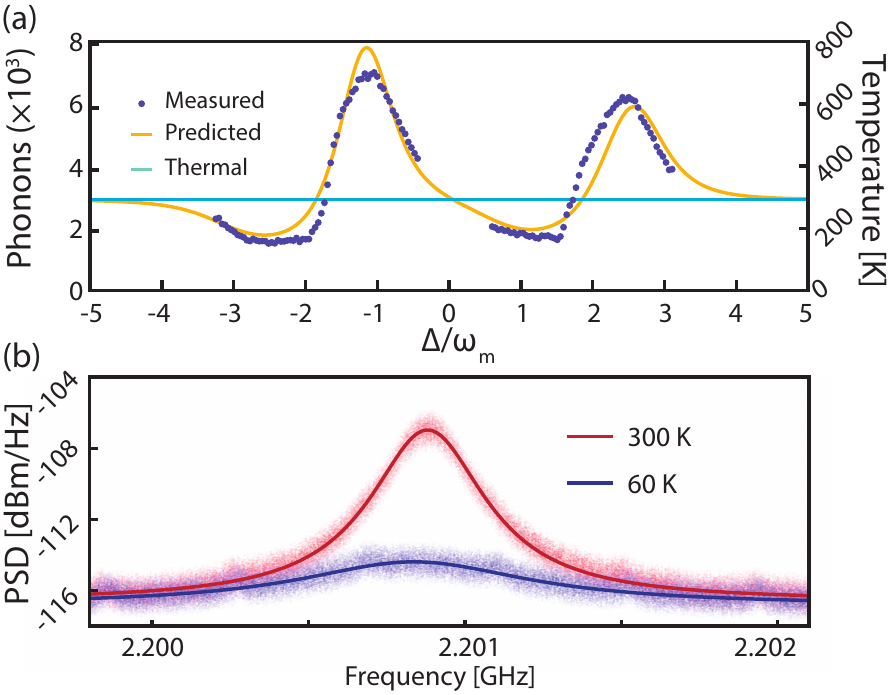, width=1\linewidth}
 \caption{Optomechanical heating and cooling. (a) Measured phonon occupation and temperature of  the RBM, and the corresponding values predicted from the fit to $\Gamma_\text{opt}$ in Fig.\ \ref{fig:Spring}(b). (b) PSD at the operating point $\Delta\sim -\kappa_\text{bs}/2-\omega_\text{m}$ of maxiumum cooling (blue), and at $\Delta$ tuned to the point of zero damping (red). Small peaks at intervals of 500 kHz are from technical noise.}
\label{fig:Gamma}
\end{figure}

\section{Optomechanically induced transparency}
Optomechanically induced transparency (OMIT) is a signature of coherent coupling between optical and mechanical resonances, and has been demonstrated in cavity optomechanical systems such as microtoroids \cite{ref:weis2010oit}, optomechanical crystals (OMC's) \cite{ref:safavi2011eit}, and microdisks \cite{ref:liu2013eit}. OMIT occurs when a strong control field ($\omega_\text{c}$) is red--detuned from the microdisk such that $\Delta_\text{oc} = \omega_\text{o}-\omega_\text{c} = \omega_\text{m}$, resulting in destructive interference between anti-Stokes photons scattered from the control field and a weak probe field ($\omega_\text{p}$). This creates a transparency window (dip) in the probe transmission (reflection) spectrum when $\Delta_\text{pc} = \omega_\text{p}-\omega_\text{o} = \omega_\text{m}$ (corresponding to $\omega_\text{p} = \omega_\text{o}$ if the control field detuning condition is ideally satisfied) whose amplitude and width depends on the cooperativity \cite{ref:aspelmeyer2014co, ref:weis2010oit, ref:safavi2011eit}.

\begin{figure}[ht]
\epsfig{figure=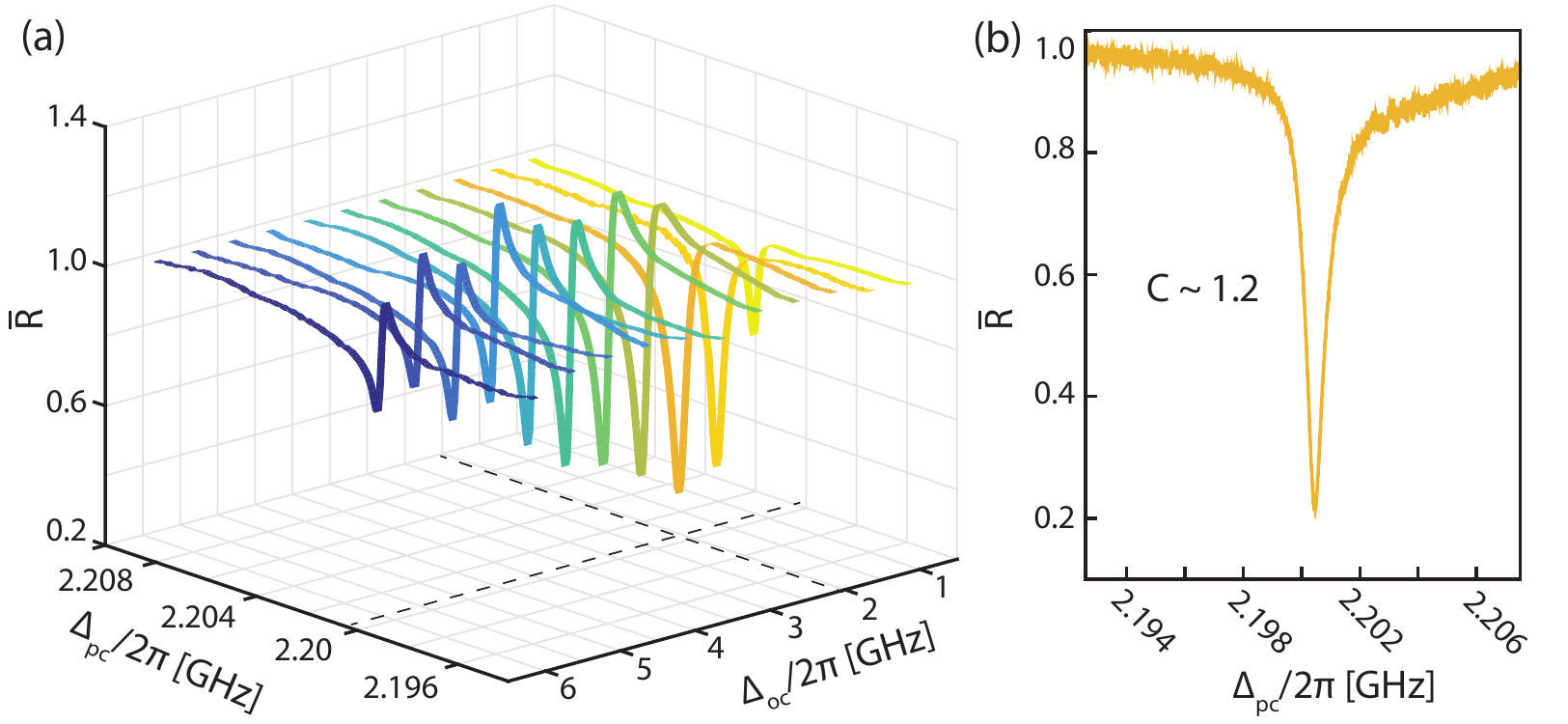, width=1\linewidth}
 \caption{Optomechanically induced transparency. (a) Probe reflection,$\bar{R}$, as a function of $\Delta_\text{oc}$ and $\Delta_\text{pc}$ shown for OMIT using the first mode of the doublet. (b) OMIT response corresponding to $C \sim 1.2$ for $N \sim 2.7\times10^6$, with $\Delta_\text{pc}$ = $\Delta_\text{oc}$ = $\omega_\text{m}$.}
\label{fig:OMIT}
\end{figure}

To characterize OMIT in the diamond microdisks, the laser output was amplified to $P_i \sim 40~\text{mW}$, and its wavelength was slowly stepped across the cavity resonance, creating a control field with varying $\Delta_\text{oc}$. At each $\Delta_\text{oc}$, a phase electro--optic modulator (EOM) driven by a vector network analyzer (VNA:Keysight E5036A) was used to create a sideband on the control field that serves as the probe, and whose frequency can be swept accross the cavity resonance, varying $\Delta_\text{pc}$. The probe field reflected by the microdisk back into the fiber taper was measured using a high--bandwidth photoreceiver connected to an optical circulator, and analyzed by the VNA.   The low-frequency mode of the microdisk doublet $(\omega_\text{o} - \frac{\kappa_\text{bs}}{2})$ was used for all of the measurements described below. Figure \ref{fig:OMIT}(a) shows the results of these measurements for several $\Delta_\text{oc}$, with each exhibiting a sharp OMIT feature when $\Delta_\text{pc} = \omega_\text{m}$. 	Here $\overline{R}$ is the reflectivity normalized by its maximum value in absence of OMIT. When $\Delta_\text{oc}$ is tuned away from $\omega_\text{m}$ the OMIT feature exhibits a Fano shape due to the phase difference between the scattered control field and the probe field. At the OMIT condition $\Delta_\text{oc} = \omega_\text{m}$ the dip amplitude reaches a maximum, as shown in detail in Fig.\ \ref{fig:OMIT}(b). From the dip amplitude $\sim 0.8 = 1- 1/(1+C)^2$ \cite{ref:liu2013eit}, cooperativity $C \sim 1.2$ was extracted. This $C$ was achieved with an intracavity photon number of $N \sim 2.7\times10^6$, and corresponds to   $g_0/2\pi =18~\text{kHz}$, in excellent agreement with the value predicted from the optomechanical spring effect fits in Figs.\ \ref{fig:Gamma}(a) and (b).

Compared with previous demonstrations of OMIT in microdisks \cite{ref:liu2013eit}, the diamond devices demonstrated here support nearly two orders of magnitude higher $N$ and 2.6 times higher $C$, despite their lower $Q_\text{o}$. Furthermore, this performance is realized without thermal stabilization or cryogenic cooling. This enables optomechanical cooling with large $N$ in the red-detuned regime not accessible in previous studies \cite{ref:mitchell2016scd}. Additionally, the  microdisk geometry combined with the broadband transparency of diamond allows these devices to simultaneously support high-$Q_\text{o}$ optical modes spanning a broad wavelength range, for example at both the 637 nm range of diamond NV centre emission and in the 1550 nm telecommunications wavelength band\cite{ref:mitchell2016scd}. The $C>1$ OMIT shown here will allows these multiwavelength cavities to be used for optomechanical wavelength conversion \cite{ref:liu2013eit, ref:hill2012cow}, and the broad microdisk mode spectrum will allow conversion over a larger range than diamond OMC \cite{ref:burek2016doc}. Finally, any improvement to $Q_\text{m}$, either through low-temperature operation \cite{ref:khanaliloo2015dnw} or engineering of the microdisk connection to the pedestal \cite{ref:anetsberger2008dul} would greatly increase the maximum achievable cooperativity.

\section{Conclusion}
In summary we have demonstrated optomechanically induced transparency, cooling, and full characterization of the optical spring effect in single-crystal diamond microdisks. Access to the red--detuned sideband, and the associated aforementioned phenomena was enabled by improving the thermal stability of the microdisk by modify the pedestal geometry.  Future work will seek to further enhance $Q_\text{o}$ and $Q_\text{m}$ by improving the fabrication process and investigating post-fabrication surface treatments. Furthermore, recent work has demonstrated that the diamond undercutting fabrication technique utilized here may be applied to other geometries such as photonic crystals \cite{ref:mouradian2017rpc}, which holds promise for the fabrication of future optomechanical devices. Finally, implantation of NV's or SiV's in these devices will allow the study of the interaction of solid state qubits with coherently driven cavity--optomechanics, as spin-photon coupling rates between the RBM and NV ground state are predicted to reach 0.6 MHz \cite{ref:mitchell2016scd}.

\section*{Funding}
This work was supported by NRC, CFI, iCORE/AITF, and NSERC.

\input{manuscript_bib.bbl}

\clearpage

\onecolumngrid

\setcounter{equation}{0}
\setcounter{figure}{0}
\setcounter{section}{0}
\setcounter{subsection}{0}
\setcounter{table}{0}
\setcounter{page}{1}
\makeatletter
\renewcommand{\theequation}{S\arabic{equation}}
\renewcommand{\thetable}{S\arabic{table}}
\renewcommand{\thefigure}{S\arabic{figure}}
\renewcommand{ \citenumfont}[1]{S#1}
\renewcommand{\bibnumfmt}[1]{[S#1]}

\section*{Supplementary Information}

\noindent This document provides supplementary information to ``Optomechanically induced transparency and cooling in thermally stable diamond microcavities''. First we discuss the modified fabrication process utilized in this work. Secondly we discuss the thermal response of our cavity including measuring the thermal time constant and central wavelength dependence on temperature. We then discuss modelling the optical spring effect, and transduction for an optical doublet.

\subsection{Modified fabrication process}

\begin{figure}[h]
\epsfig{figure=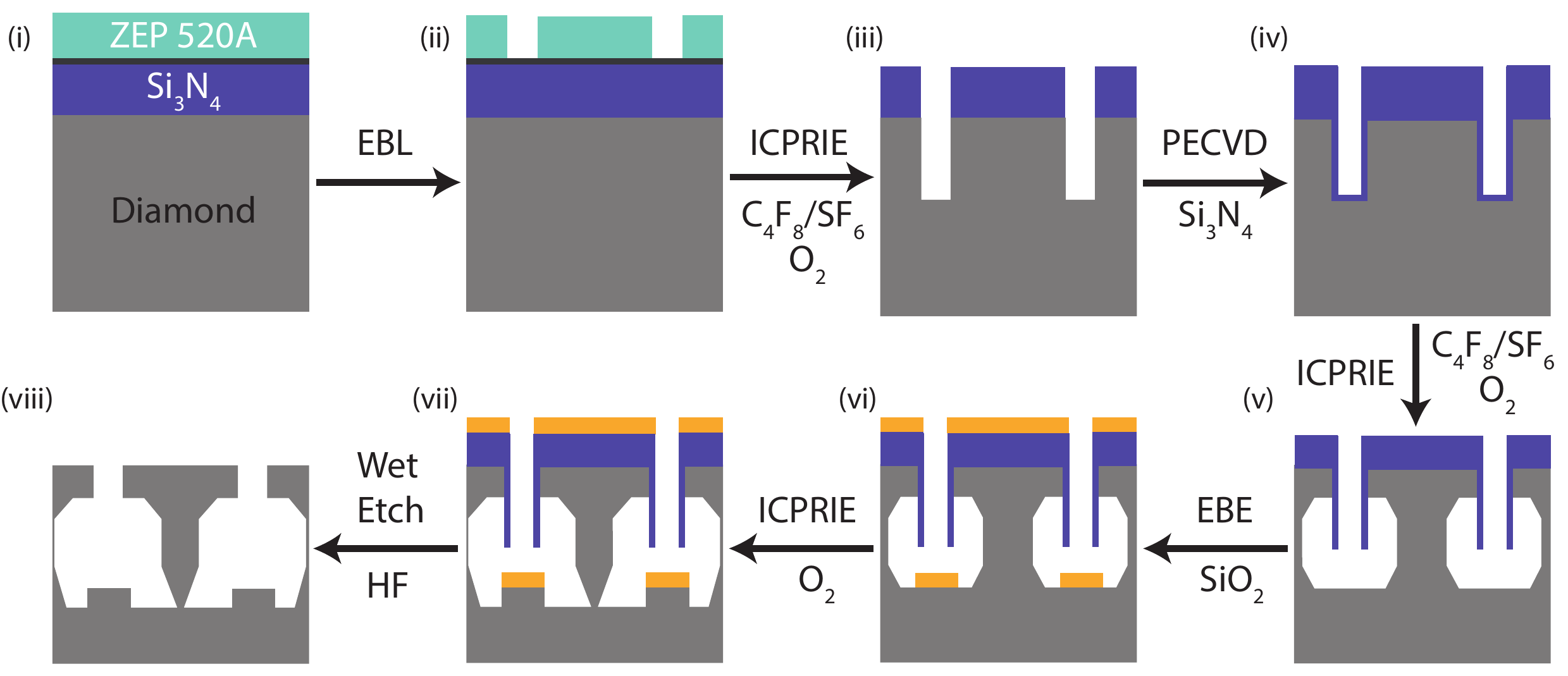, width=1\linewidth}
 \caption{Modified plasma undercutting method. (i) Polished single--crystal diamond chips purchased from element 6 are cleaned in boiling piranha and coated with a 300 nm thick PECVD Si$_3$N$_4$ layer, coated with $\sim 5$ nm of Ti as an anticharging layer, and EBL resist (ZEP 520A). (ii) The microdisks are patterned in ZEP using EBL and developed in ZED N50. (iii) Patterns are transferred to the Si$_3$N$_4$ hard mask using an ICPRIE etch. ZEP is removed using a deep-UV exposure and Remover PG. (iv) Patterns are transferred to the diamond using an anisotropic O2 plasma ICPRIE etch, followed by sidewall protection via a conformal coating of PECVD Si$_3$N$_4$. (v) A short ICPRIE etch removes Si$_3$N$_4$ from the bottom of the etch windows, followed by an initial zero bias O$_2$ ICPRIE plasma which partially undercuts the microdisks. (vi) A $\sim 100$ nm layer of SiO$_2$ is deposited via electron beam evaporation (EBE). (vii) A second zero bias O$_2$ plasma etch is performed, finishing the plasma undercutting process. (viii) The sample is soaked in HF to remove the remaining Si$_3$N$_4$ layer, followed by a piranha clean.}
\label{fig:Fabrication}
\end{figure}

Compared to previous work fabricating microdisks from bulk single--crystal diamond a modified undercutting approach was utilized in this work, as described in Fig.\ \ref{fig:Fabrication}.

\subsection{Thermal response}

Following the analysis of Carmon et al \cite{ref:supp_carmon2004dtb}, we can write an equation governing the time evolution of the cavity temperature in response to an input field with power $P_\text{in}(t)$,
\begin{equation}\label{eq:temperatureDE}
\frac{d}{dt} \Delta T = P_\text{in}(t) f\left(\lambda_p,\Delta T\right) - \frac{\Delta T}{\tau_\text{th}},
\end{equation}
where for a singlet mode,
\begin{equation}
f\left(\lambda,\Delta T\right) = \frac{\eta Q_\text{o}}{C_p Q_\text{abs}} \frac{ \left[\Delta \lambda /2 \right]^2}{\left[\lambda_p-\lambda_o(1+\beta\Delta T)\right]^2+\left[\Delta \lambda /2\right]^2}.
\end{equation}
In the above $\Delta T$ is the difference between the cavity temperature and environment temperature, $\tau_\text{th}$ is the thermal time constant, $C_p$ is the heat capacity of the cavity, and $\beta$ is defined as a temperature coefficient of resonance--wavelength accounting for thermal expansion and refractive index pertubations, $\beta = \epsilon +\frac{dn}{dT}/n_0$. The fraction of light coupled into the cavity is defined as $\eta$, $\lambda_p$ is the input field wavelength, $\lambda_o(\Delta T)$ is the cavity resonance wavelength, $Q_\text{o}$ is the cavity intrinsic quality factor, and $Q_\text{abs}$ is the quality factor due to absorption only.

\begin{figure}[h]
\epsfig{figure=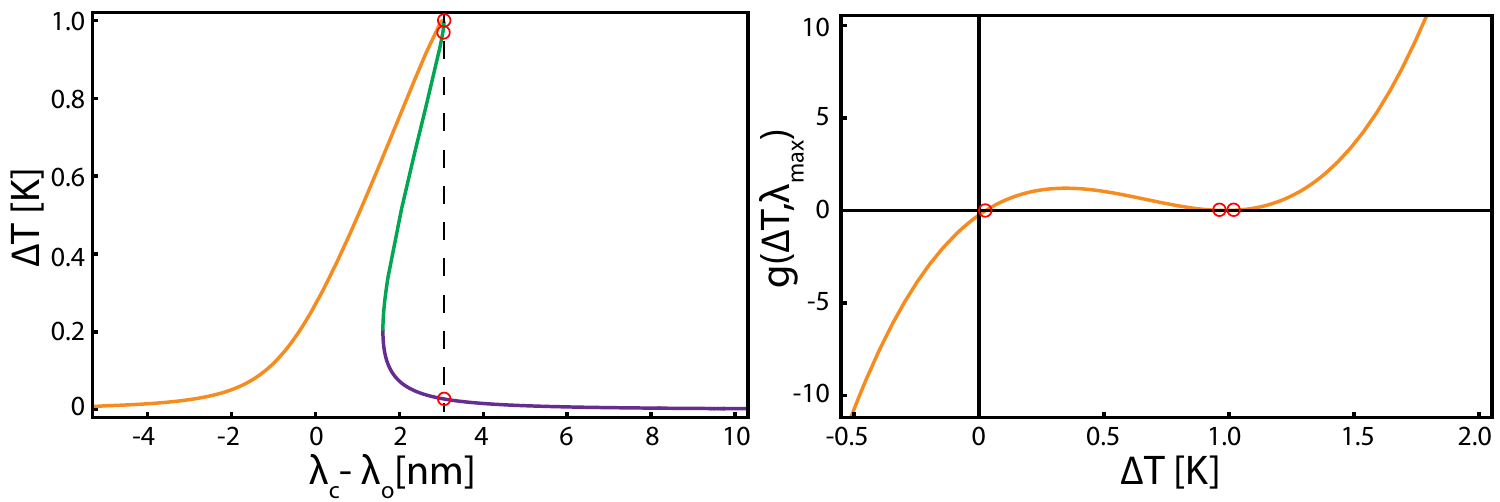, width=1\linewidth}
 \caption{(a)Cavity temperature at equilibrium as a function of laser--cavity detuning found by numerically solving Eqn. \ref{eq:temperatureDE}, with $\frac{d}{dt} \Delta T = 0$. (b) Equilibrium temperature solutions at maximum shift in resonance wavelength, $\lambda_\text{max}$.}
\label{fig:Bistable}
\end{figure}

For a constant input signal, the equilibrium temperature $\Delta T_\text{eq}$ of the cavity can be found by setting $\frac{d}{dt} \Delta T = 0$, and solving the cubic equation $a (\Delta T_\text{eq})^3+b (\Delta T_\text{eq})^2+c (\Delta T_\text{eq})+d = 0$. As only real solutions to this equation are physical, we can deduce the number of valid solutions at equilibrium by determining the number of real roots associated with the cubic equation. To do so one can evaluate the discriminant, $\Delta$, as,
\begin{align}
\Delta &= \frac{q^2}{4}+\frac{p^3}{27}, \\
p & = \frac{3ac-b^2}{3a^2}, \\
q &= \frac{2b^3-9abc+27a^2d}{27a^3}.
\end{align}
This can be divided into three distinct cases,
\begin{align}
\Delta < 0: & \,\text{Three real, distinct roots.} \\\nonumber
\Delta = 0: & \, \text{Three real, degenerate roots.} \\\nonumber
\Delta > 0: & \, \text{One real root, and two complex roots.}
\end{align}
The case where $\Delta <0$ is shown in Fig.\ \ref{fig:Bistable}, where 3 real distinct roots exist resulting in a thermal bistability.

To determine if the red-detuned side of the cavity is accessible for a given input pump power, we evaluate $\Delta$ at zero detuning, e.g. $\lambda_p=\lambda_o(1+\beta\Delta T)$. This leads to the requirement for access to red detunings,
\begin{equation}
\frac{1}{\tau_\text{th} Q_\text{o}} > \beta \left(\frac{\eta Q_\text{o}}{Q_{abs}}\right) \left(\frac{P_\text{in}}{C_p}\right).
\end{equation}

\begin{figure}[h]
\epsfig{figure=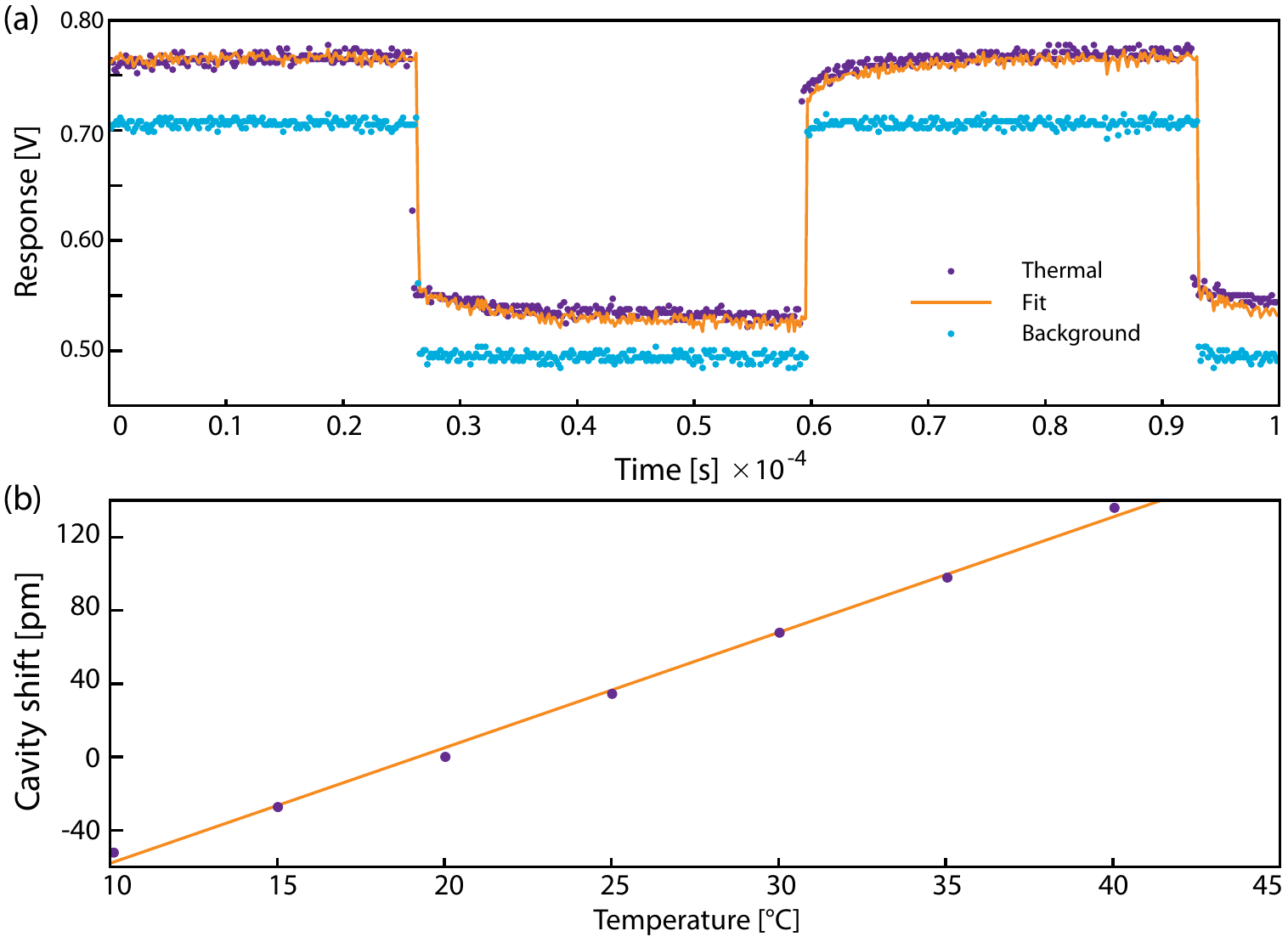, width=1\linewidth}
 \caption{(a) Transmission of an optical cavity in response to a square wave driving function. (b) Cavity shift, relative to ambient conditions, as a function of temperature for the device and mode used in the main text. The cavity temperature was varied using a thermoelectric heater. The linear line of best fit shown was used to calculate the laser absorption induced heating present in the cooling experiment described in the main text.}
\label{fig:Thermal}
\end{figure}

To measure the thermal time constant, the input field is modulated between two distinct powers, as shown in Fig.\ \ref{fig:Thermal}(a). For relatively small modulations, the time evolution of the cavity temperature will resemble the input field, low pass filtered by the finite response time of the thermal cavity shift. This can then be read out by choosing a laser drive wavelength where changes in the cavity resonance frequency result in a change in optical transmission, namely,
\begin{equation}
P_\text{out}(t) \approx T\left[1 + \frac{\partial T}{\partial \lambda_o} \frac{\partial \lambda_o}{\partial \Delta T} \Delta T\right] P_\text{in}(t),
\label{eq:transmissionEQ}
\end{equation}
where $P_\text{out}$ is the power output into the taper and $T$ is the transmission through the cavity. Together Eqs. \ref{eq:temperatureDE}-\ref{eq:transmissionEQ} was used to fit experimental data to derive $\tau_\text{th}$.

In the cooling experiments described in the main text, the application of a strong pump laser caused the device to heat. The degree of this temperature shift was derived by measuring the shift in $\lambda_o$ as a function of temperature in a separate experiment, as shown in Fig.\ \ref{fig:Thermal}(b).

\subsection{Optical spring effect}

Microdisk resonators typically support degenerate clockwise and counter clockwise propagating modes, with amplitudes $\alpha_\text{cw}, \alpha_{ccw}$ respectively. In the case that the coupling between these modes due to backscattering exceeds loss to all other channels, standing wave modes are supported, which are either a symmetric $\alpha_\text{s}=1/\sqrt{2}\left(\alpha_\text{cw}+ \alpha_\text{ccw} \right)$ or anti-symmetric $\alpha_\text{a}=1/\sqrt{2}\left(\alpha_\text{cw} - \alpha_\text{ccw} \right)$ combinations of the travelling wave modes \cite{ref:supp_borselli2006hqm}.

Using the fact that $\langle\alpha_a | \alpha_s \rangle=0$, we can write the optomechanical spring effect as \cite{ref:supp_aspelmeyer2014co},
\begin{align}
\delta \omega_m =& g_\text{s}^2 \left[\frac{\Delta-\omega_m}{\kappa_s^2/4+(\Delta-\kappa_\text{bs}/2-\omega_m)^2} \right] \nonumber\\
&+g_\text{s}^2 \left[\frac{\Delta+\omega_m}{\kappa_s^2/4+(\Delta-\kappa_\text{bs}/2-\omega_m)^2} \right] \nonumber\\
&+g_\text{a}^2 \left[\frac{\Delta-\omega_m}{\kappa_a^2/4+(\Delta+\kappa_\text{bs}/2-\omega_m)^2} \right] \nonumber\\
&+g_\text{a}^2 \left[\frac{\Delta+\omega_m}{\kappa_a^2/4+(\Delta+\kappa_\text{bs}/2-\omega_m)^2} \right]
\end{align}
where $g_a,g_s$ are the optomechanical coupling parameters to the symmetric and the anti-symmetric modes, $\kappa_a,\kappa_s$ are the cavity decay rates, $\kappa_{bs}$ is the backscattering coupling rate, $\omega_m$ is the mechanical frequency rate and $\Delta$ is the laser-cavity detuning rate.

Using similar arguments, it can be shown that the optomechanical damping rate for doublets is,
\begin{align}
\Gamma_\text{opt} =& g_\text{s}^2 \left[\frac{\kappa_s}{\kappa_s^2/4+(\Delta-\kappa_\text{bs}/2-\omega_m)^2} \right] \nonumber\\
&-g_\text{s}^2 \left[\frac{\kappa_s}{\kappa_s^2/4+(\Delta-\kappa_\text{bs}/2-\omega_m)^2} \right] \nonumber\\
&+g_\text{a}^2 \left[\frac{\kappa_a}{\kappa_a^2/4+(\Delta+\kappa_\text{bs}/2-\omega_m)^2} \right] \nonumber\\
&-g_\text{a}^2 \left[\frac{\kappa_a}{\kappa_a^2/4+(\Delta+\kappa_\text{bs}/2-\omega_m)^2} \right].
\end{align}

\subsection{Transduction}

Modelling the mechanical mode amplitude as $x(t)=x_0 \cos \left( \omega_m t \right)$, where $x_0$ is the zero point fluctuation of the mechanical mode, the amplitudes of the field in the symmetric optical mode $\alpha_\text{s}$ and anti-symmetric optical mode $\alpha_\text{a}$ are modulated by the mechanics as:
\begin{equation}
\alpha_\text{s,a}(t) = \alpha_\text{in} \sqrt{\frac{\kappa_\text{ex}}{2}} \mathcal{L}_\text{s,a}(0) \times \left[1 - \frac{ \text{i}x_0 G_\text{s,a} \mathcal{L}_\text{s,a}( \omega_m) }{2} e^{-\text{i}\omega_m t}
- \frac{ \text{i}x_0 G_\text{s,a} \mathcal{L}_\text{s,a}(-\omega_m) }{2} e^{ \text{i}\omega_m t} \right]
\end{equation}
where $\kappa_\text{ex}$ is the external coupling rate to the fiber, $ \left| \alpha_\text{in} \right|^2 = P_\text{in}$, and
\begin{equation}
\mathcal{L}_\text{s,a}( \omega) = \frac{1}{-\text{i}\left(\omega \pm \kappa_\text{bs}/2+\Delta \right)+\kappa_\text{s,a}/2}.
\end{equation}

In the experiment we measure the reflected signal, R, which may be written in terms of the optical mode amplitudes as,
\begin{equation}
R=\sqrt{\frac{\kappa_\text{ex}}{2}} \left|\alpha_\text{s}-\alpha_\text{a}\right|^2 P_\text{in}.
\end{equation}

Solving for the power spectral density of the reflected signal, $S_\text{RR}$ in terms of the power spectral density of the driven harmonic mechanics signal, $S_\text{xx}$ and ignoring small terms, we can find the transduction coefficients $K(\Delta)$ which satisfies the expression:

\begin{equation}
\frac{S_\text{RR}(\omega)}{P_\text{in}^2} = \left| K(\Delta) \right|^2 S_{xx}.
\end{equation}
where,
\begin{equation}
S_{xx} = 2 \pi x_0^2 \left[\delta(\omega-\omega_m)+\delta(\omega+\omega_m)\right]
\end{equation}

\end{document}

%% file: manuscript_bib.bbl
%